\documentclass[aps, onecolumn, floatfix, preprintnumbers, superscriptaddress, nofootinbib, 11pt]{revtex4-2}

\usepackage{setspace}
\usepackage{graphics,amssymb,amsmath,epsfig,color}
\usepackage{graphicx}
\usepackage{graphicx, color, xcolor, soul}

\usepackage[raggedright, small]{titlesec}

\titlespacing*{\section}
  {0pt}{2.0ex plus 1ex minus .2ex}{0.8ex plus .2ex} 

\titlespacing*{\subsection}
  {0pt}{1.5ex plus 1ex minus .2ex}{0.6ex plus .2ex}
\makeatletter

\def\@hangfrom@section#1#2#3{\@hangfrom{#1#2}#3}
\def\@hangfroms@section#1#2{#1#2}

\begin{document}

\def\scititle{
	Ultra-stable laser with sub-Hz intrinsic linewidth using a silicon photonic chip
}


\title{\bfseries \boldmath \large \scititle}
\author{Mohamad Hossein Idjadi$^{\dagger,*}$}
\affiliation{Nokia Bell Labs, 600 Mountain Ave., Murray Hill, NJ 07974, USA.}

\author{Haotian Cheng$^{\dagger}$}
\affiliation{Department of Applied Physics, Yale University, New Haven, CT 06511, USA}

\author{Benjia Li$^{\dagger}$}
\affiliation{Department of Applied Physics, Yale University, New Haven, CT 06511, USA}

\author{Tzu-Yung Huang}
\affiliation{Nokia Bell Labs, 600 Mountain Ave., Murray Hill, NJ 07974, USA.}

\author{Kwangwoong Kim}
\affiliation{Nokia Bell Labs, 600 Mountain Ave., Murray Hill, NJ 07974, USA.}

\author{Naijun Jin}
\affiliation{Department of Applied Physics, Yale University, New Haven, CT 06511, USA}


\author{Peter T. Rakich$^{\ddagger}$}
\affiliation{Department of Applied Physics, Yale University, New Haven, CT 06511, USA}
\begin{abstract}
Here, we demonstrate a novel optical self-injection-locking scheme integrated with a modulation-free laser frequency stabilization technique co-integrated with a compact vacuum-gap micro-Fabry-P\'erot cavity ($\mu$FP), with Q-factor of $3.46\times10^9$, to realize an ultra-stable laser with sub-Hz fundamental linewidth using a silicon photonic chip. The demonstrated electro-optic self-injection locking (EOSIL) chip suppresses a commercial DFB diode laser’s frequency noise power spectral density by more than 110~dB at 1~Hz offset, approaching the theoretical thermal noise limit of the $\mu$FP cavity, and enables a laser fundamental linewidth of 58~mHz with a fractional frequency instability of $8.68\times10^{-13}$ at 1~sec averaging time. The chip is fabricated using the Advanced Micro Foundry standard silicon photonic process and occupies a footprint of 10~mm$^2$. 
\end{abstract}

\maketitle
\let\thefootnote\relax\footnotetext{$^{\dagger}$These authors contributed equally.\\ $^{*}$Corresponding author: mohamad.idjadi@nokia-bell-labs.com \\$^{\ddagger}$Corresponding author: peter.rakich@yale.edu}
\section*{Introduction}
Ultra-stable lasers with sub-Hz linewidth and high fractional frequency stability are a vital part of optical systems in many demanding applications such as optical atomic clocks~\cite{udem2002optical,ludlow2015optical,jiang2011making}, photonic-assisted low-noise microwave and millimeter-wave synthesis~\cite{li2013microwave,sun2024integrated,he2024chip,kudelin2024photonic,cheng2025spiral}, precision spectroscopy~\cite{demtroder1973laser}, coherent optical communication~\cite{fang2024overcoming} and LiDAR~\cite{lihachev2022low}, quantum technology~\cite{caron2026single, xu2019sensing}, and seismic sensing~\cite{marra2018ultrastable} where laser phase noise and long-term stability are usually the bottleneck of the ultimate system performance.

The state of the art in laser stability is achieved by stabilizing a laser to a high-finesse Fabry-P\'erot cavity made from ultra-low-expansion (ULE) glass as a spacer held inside a vacuum enclosure, typically using Pound-Drever-Hall (PDH) locking~\cite{drever1983laser}. While these systems reach exceptional performance, they rely on bulk optics, free-space alignment, and discrete electronics, which confine them to table-scale laboratory setups and limit their scalability and their deployment outside controlled environments.

Considerable effort has therefore been directed toward miniaturizing both the laser and the frequency reference using integrated photonics~\cite{liang2015ultralow,liu202236,xiang20233d,heim2025hybrid,liu2025large}. Self-injection locking (SIL) to high-quality-factor integrated resonators has been shown to be an effective technique to narrow the linewidth of chip-scale lasers, passively suppressing frequency noise over a wide bandwidth without active
electronic feedback~\cite{liang2015ultralow,xiang20233d}. However, SIL alone does not guarantee long-term frequency stability~\cite{cheng2025harnessing}. Active
electro-optic (EO) stabilization will ensure active adjustment of laser frequency and guarantees stability over long-term. The conventional PDH approach, one of the most widely used EO laser locking methods, requires an optical phase modulator and a demodulation chain, and its performance is ultimately bounded by residual amplitude modulation
(RAM)~\cite{shi2018suppression}.

Here, we address these limitations with an electro-optic self-injection-locking (EOSIL) architecture that co-integrates SIL with a modulation-free, cavity-coupled Mach Zehnder interferometer (CCMZI) electro-optic stabilization loop on a single foundry-fabricated silicon photonic chip~\cite{idjadi2024modulation,idjadi2026chip}, referenced to a compact vacuum-gap micro-Fabry-P\'erot ($\mu$FP) cavity~\cite{jin2022micro, cheng2023novel,liu2024ultrastable}. Self-injection locking passively
suppresses the broadband frequency noise, while the CCMZI loop derives a modulation-free error signal directly from the $\mu$FP resonance to actively correct laser frequency and suppress its phase noise at low-offset and ensure long-term stability. Using this architecture, the EOSIL chip suppresses the frequency noise power spectral density of a commercial DFB diode laser by more than 110~dB at 1~Hz offset, approaching the thermal noise limit of the $\mu$FP cavity, and reduces the integral ($\beta$-separation line method~\cite{di2010simple}) and fundamental linewidths from 3.8~MHz and 81.4~kHz to 38~Hz and 58~mHz, respectively, while reaching a fractional frequency instability of $8.68\times10^{-13}$ at 1~sec averaging time. The photonic chip is fabricated in the Advanced Micro Foundry (AMF) standard silicon-on-insulator process within a 10~mm$^2$ footprint. The proposed EOSIL architecture, implemented on a silicon chip, paves the way toward scalable and deployable ultra-stable laser sources.
\section*{Results}
\subsection*{Principle of operation}
\begin{figure}[tb]
    \centering
    \includegraphics[width=\textwidth]{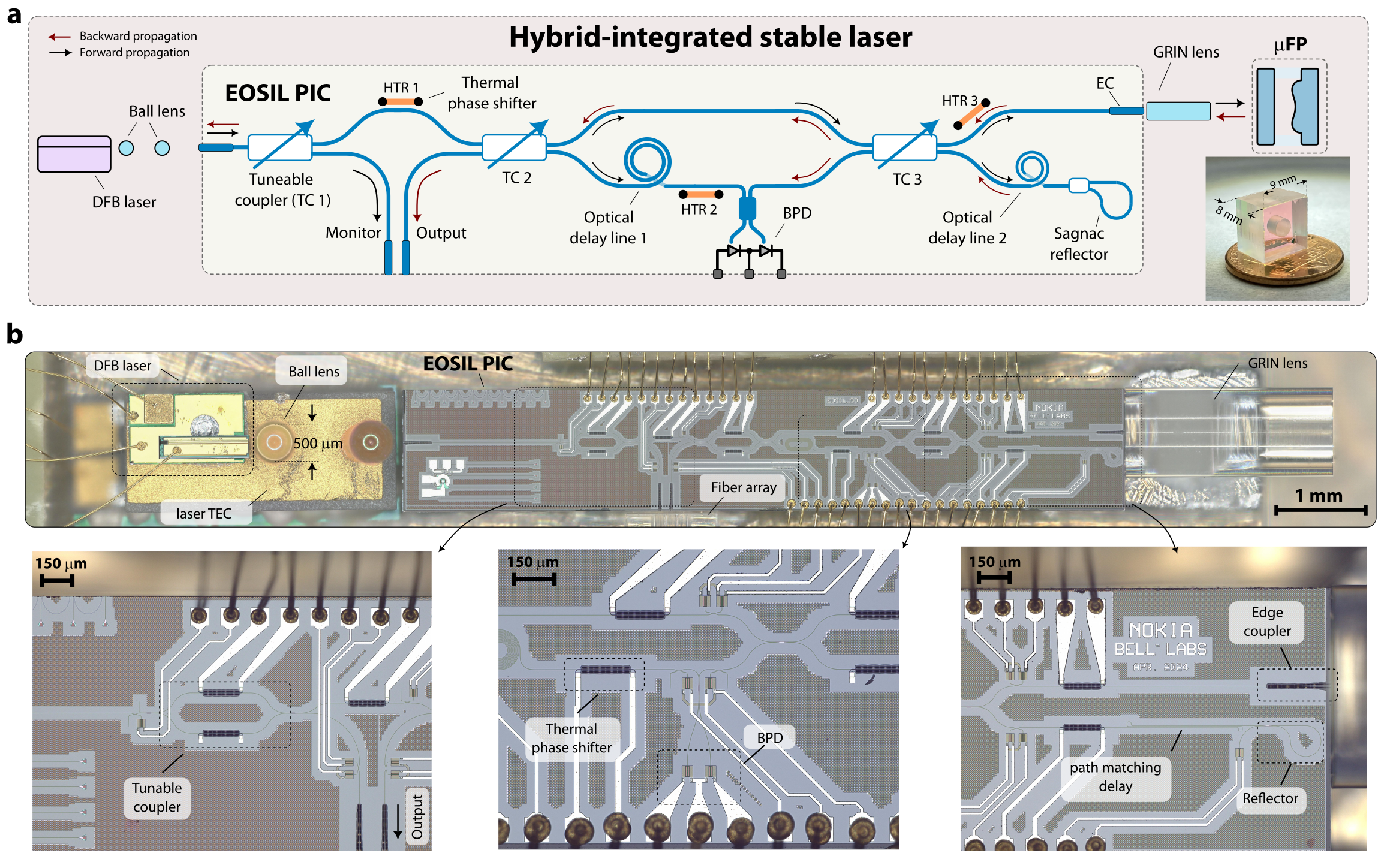}
    \caption{\textbf{Electro-optic self-injection locking (EOSIL) laser stabilization.} \textbf{a,} Block diagram of the EOSIL photonic chip. The vacuum-gap micro-Fabry–P\'erot ($\mu$FP) cavity is made with ultra-low-expansion (ULE) spacers with a thickness of 4 mm, corresponding to a free spectral range of 37 GHz. The cavity has a measured Q-factor of $3.46\times10^9$ at frequency of 192.94~ THz. The diced cavity dimensions are 9 mm $\times$ 9 mm $\times$ 8 mm. \textbf{b,} Microphotograph of the implemented EOSIL photonic integrated circuit (PIC), fabricated using the Advanced Micro Foundry (AMF) standard silicon-on-insulator (SOI) photonic process. }
    \label{fig1_pic}
\end{figure}

Figure~\ref{fig1_pic}a shows a block diagram of the hybrid-integrated laser, which
combines self-injection locking (SIL) with a cavity-coupled Mach-Zehnder
interferometer (CCMZI) electro-optic stabilization architecture. The EOSIL silicon
photonic integrated circuit (PIC) serves as the interface between the DFB laser diode chip and the vacuum-gap micro Fabry-P\'erot ($\mu$FP) cavity. The EOSIL PIC comprises two interferometers: a Michelson interferometer that generates the optical feedback returned to the DFB laser diode, and a Mach-Zehnder interferometer that generates an electronic error signal for long-term frequency stabilization.

Light from the DFB laser is coupled into the chip through a ball lens and split by a tunable coupler (TC~1). One output is routed to the first channels of a 1$\times$2 fiber array and serves as the laser output monitoring, while the other propagates through a thermal phase shifter (HTR~1) followed by TC~2 at the input of the MZI. HTR~1 controls the phase of the optical feedback signal. The upper MZI arm feeds the 2$\times$2 tunable coupler TC~3, which defines the two reflective paths of the Michelson interferometer. Its top port is routed to an edge coupler and, via a gradient-index (GRIN) lens for mode size matching, to the $\mu$FP cavity, from which the reflected field is collected. The bottom port forms a reference path in which the light is fully reflected by a Sagnac loop mirror, with optical delay line~2 matching the optical path length of the top arm. The two reflections recombine at TC~3, and its coupling ratio is tuned so that the interfering fields have equal amplitude at the point of interference. HTR~3 sets the phase such that the power returned to the DFB laser is maximized on the $\mu$FP resonance, establishing the self-injection-locking condition~\cite{cheng2025harnessing}. The lower arm of the MZI provides the reference for generation of an electronic error signal. It is delayed by a silicon photonic waveguide spiral (optical delay line~1) to match the optical path of the upper arm, including the round trip to the cavity. This field interferes with the light returned from TC~3 and is detected by on-chip germanium balanced photodetectors (BPD), generating the electronic error signal used for long-term frequency stabilization~\cite{idjadi2024modulation, idjadi2026chip}. Tuning HTR~2 adjusts the interferometric phase of the MZI and thereby shapes the error signal about the lock point.

Figure~\ref{fig1_pic}b shows a micrograph of the packaged EOSIL PIC, together with the DFB laser diode and the GRIN lens that couples light to the $\mu$FP cavity. The chip is fabricated in the Advanced Micro Foundry (AMF) standard silicon-on-insulator (SOI) photonic process. The close-ups below highlight the principal building blocks of the circuit such as TCs, the thermal phase shifters, the BPD, the edge coupler and Sagnac reflector. Further details of the assembly are given in the Methods.

\subsection*{EOSIL error signal and open-loop operation}
\begin{figure}[tb]
    \centering
    \includegraphics[width=\textwidth]{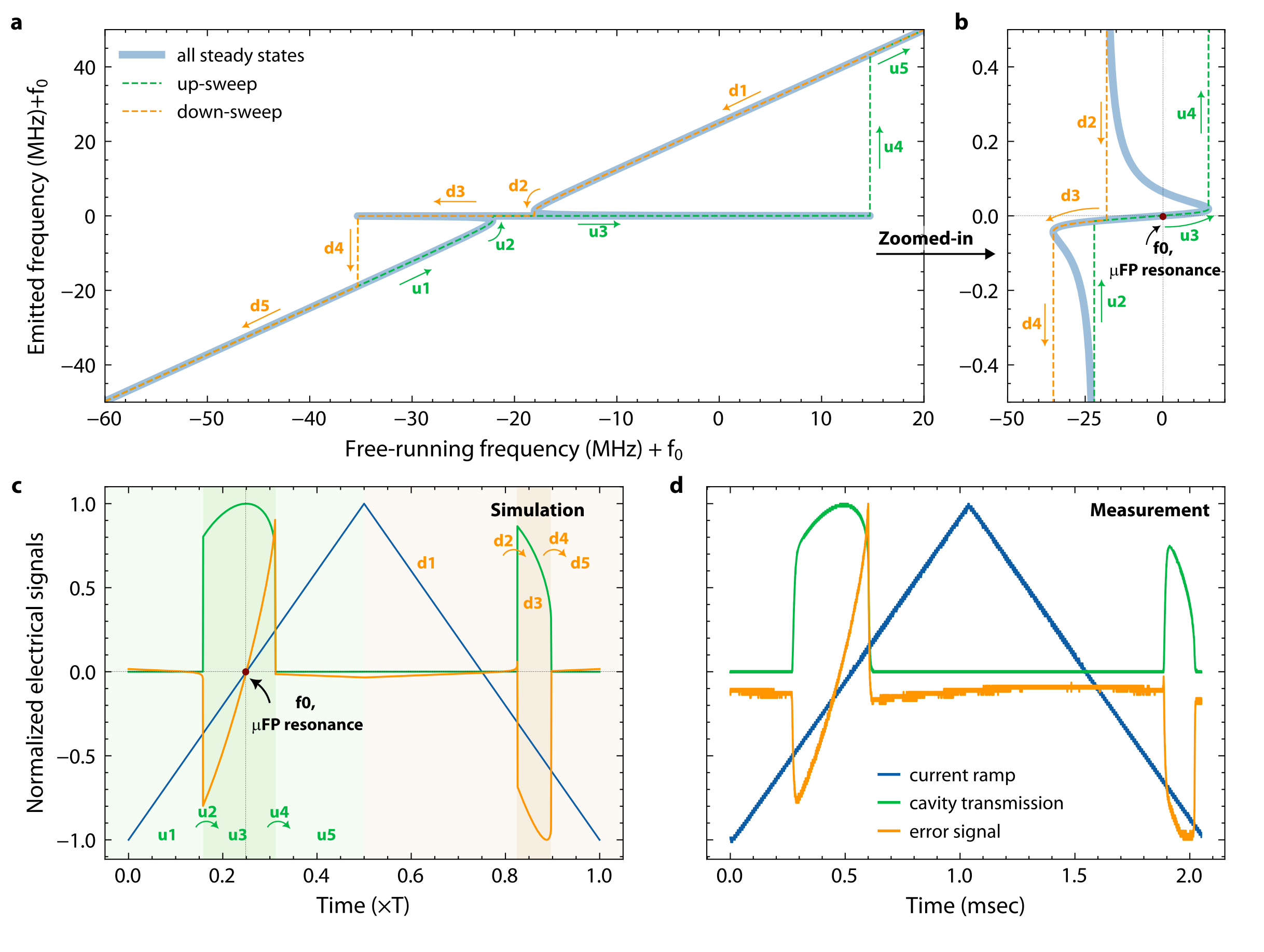}
    \caption{\textbf{Open-loop dynamics of the EOSIL laser.} \textbf{a,} Simulated self-injection-locking (SIL) curve. The emitted frequency of the laser and its free-running frequency are  both referenced to the resonance frequency of the
    $\mu$FP cavity, $f_0$. Model parameters are optimized to fit the measurement in \textbf{d}. \textbf{b,} Zoomed-in view of the boxed region in \textbf{a}, showing the locked branch and the locking point. \textbf{c,} Simulated cavity transmission (green) and electronic error signal (orange) as the bias current is ramped with a triangular waveform (blue) of period $T$. Labels correspond to the branches in \textbf{a}. \textbf{d,} Measured EOSIL open-loop response. All traces are normalized to their peak values.}
    \label{fig2_eosil_error}
\end{figure}

Figure~\ref{fig2_eosil_error} characterizes the open-loop dynamics of the EOSIL laser. Figure~\ref{fig2_eosil_error}a shows the self-injection-locking (SIL) curve, in which the emitted frequency of the laser is plotted against its free-running frequency; both are referenced to $f_0$, the resonance frequency of the $\mu$FP cavity. The steady-state behavior of the injection-locked laser is computed as the free-running frequency is swept up (green) and down (orange) by ramping the DFB laser bias current. 


On the up-sweep, the laser is far detuned from resonance and the emitted frequency follows the free-running frequency ($u1$). As the detuning approaches $f_0$, this branch of steady states terminates and the emitted frequency jumps onto the locked branch ($u2$, Fig.~\ref{fig2_eosil_error}b), where the laser is injection locked to the cavity. Within the locking range the emitted frequency remains close to $f_0$ and advances only weakly with the free-running frequency ($u3$); this strongly reduced slope reflects the frequency pulling exerted by the $\mu$FP. At the opposite edge of the locking range the locked branch reaches its turning point, where the phase condition can no longer be satisfied, and the laser jumps out of lock ($u4$) and again follows the free-running frequency ($u5$). The down-sweep traverses the same steady-state curve but visits a different set of states. Starting above resonance, the emitted frequency follows the free-running frequency ($d1$) until that branch terminates, at which point the laser captures into lock ($d2$). It then retraces the locked branch in the opposite direction ($d3$) until the branch ends and the laser unlocks ($d4$), after which it once more follows the free-running frequency ($d5$). For this set of model parameters the capture and release detunings differ between the two sweep directions, so the locked segment traversed on the down-sweep is offset from that traversed on the up-sweep. 


As described in Fig.~\ref{fig1_pic}a, the field reflected from TC~3, whose frequency is that of Fig.~\ref{fig2_eosil_error}a, interferes with the field in the lower MZI arm and is detected at the BPD to produce the electronic error signal. Applying the CCMZI transfer function~\cite{idjadi2024modulation} to the emitted frequency of Fig.~\ref{fig2_eosil_error}a yields the error signal generated by the EOSIL architecture. Figure~\ref{fig2_eosil_error}c shows the simulated error signal (orange) and cavity transmission (green) as the free-running frequency is ramped up and down by a triangular current waveform (blue) of period $T$. During $u1$ the laser is far detuned from the cavity resonance, the transmission is low, and the error signal is set entirely by the residual optical path mismatch of the MZI. The simulation here assumes that optical delay line~1 (Fig.~\ref{fig1_pic}a) does not exactly compensate the free-space path to the cavity, leaving a residual MZI fringe which
reproduces the background observed in the measurement. When the free-running frequency reaches the turning point of the SIL curve, the cavity pulls the laser onto the locked branch, and the transmission rises, peaking as the emitted frequency passes through $f_0$. Over the $u3$ interval the error signal is a steep, monotonic discriminator whose zero crossing marks the locking point and which provides the feedback signal for long-term frequency stabilization. As the free-running frequency approaches the end of the locking range the laser jumps out of lock ($u4$); the transmission collapses and, during $u5$, the error signal returns to the residual MZI response. The down-sweep reproduces the same sequence in reverse ($d1$--$d5$), with the locked interval displaced by the hysteresis of the SIL curve.

Figure~\ref{fig2_eosil_error}d shows the measured open-loop response of the EOSIL chip. A triangular waveform of 2~msec period is applied to the DFB laser bias current to chirp the DFB laser frequency. The BPD photocurrent is converted to a voltage by a transimpedance amplifier (TIA) and recorded on an oscilloscope together with the cavity transmission, which is monitored on a separate photodetector. The measured traces reproduce the sequence predicted by the simulation, including the abrupt lock and unlock transitions, the collapse of the cavity transmission outside the locking range, and the residual MZI fringe in the unlocked intervals. Given the cavity linewidth of 56~kHz, which corresponds to Q-factor of about $3.46\times10^9$ at wavelength of around 1553.8~nm, the frequency discriminator gain is about $2.8\times10^{-3}$~V Hz$^{-1}$.  

\subsection*{EOSIL closed-loop operation}
\begin{figure}[tb]
    \centering
    \includegraphics[width=\textwidth]{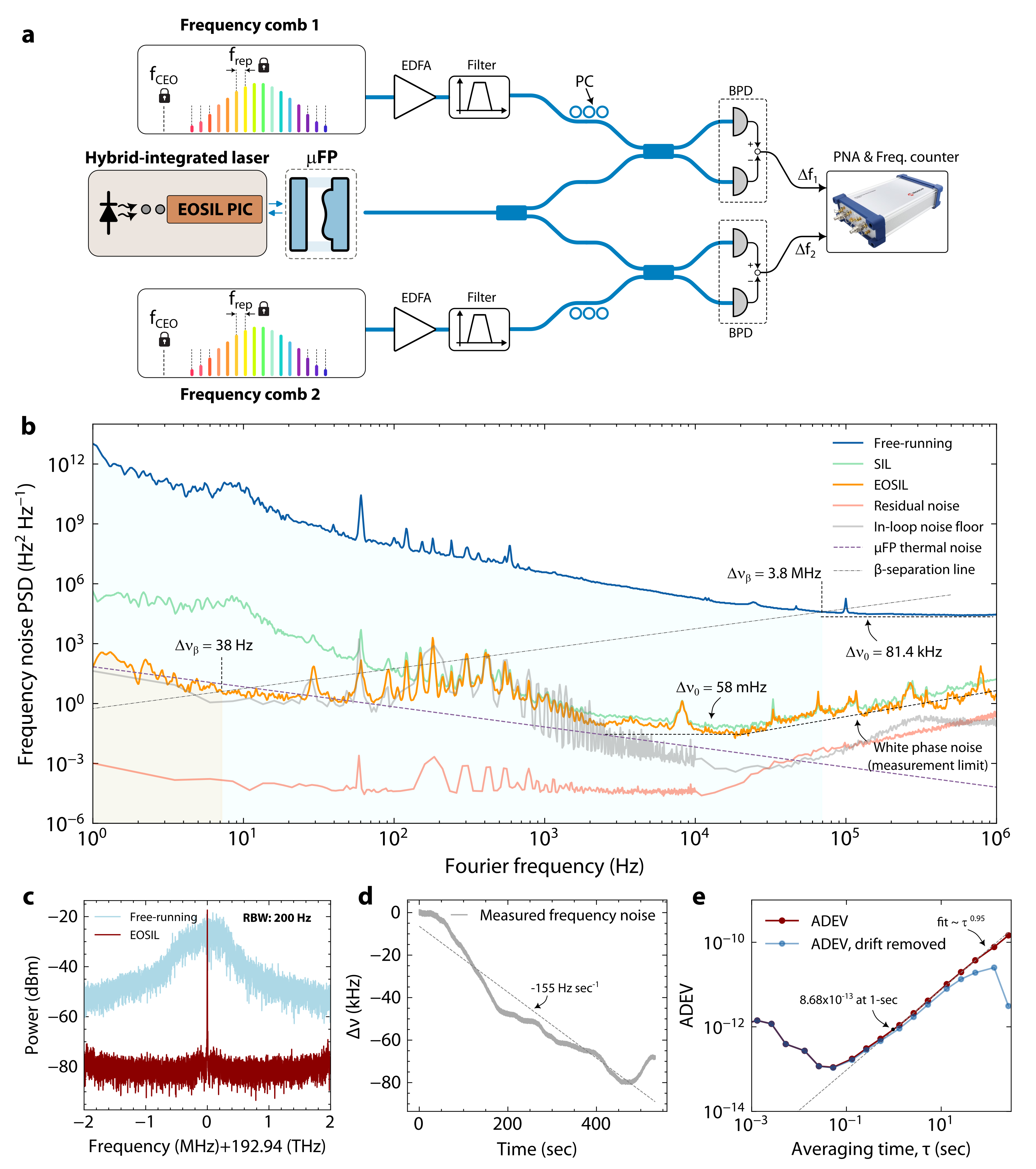}
    \caption{\textbf{EOSIL closed-loop operation.} \textbf{a}, Schematic of the frequency noise measurement setup. Each comb is amplified, spectrally filtered, polarization-matched (PC), mixed with the laser under test, and detected on a balanced photodetector (BPD) to generate two beatnotes, $\Delta f_1$ and $\Delta f_2$, recorded by a phase noise analyzer (PNA). \textbf{b}, Single-sideband frequency-noise power spectral densities (PSDs) of the laser when free-running, self-injection locked (SIL) and with electro-optic feedback engaged (EOSIL). Also shown are the residual noise, the in-loop noise floor, the $\mu$FP thermal-noise limit and the $\beta$-separation line. At high Fourier frequencies the EOSIL trace reaches the white phase noise measurement limit. \textbf{c}, RF spectrum of the beatnote $\Delta f_2$ before (free-running) and after (EOSIL) locking, measured at 200~Hz resolution bandwidth (RBW). \textbf{d},The measured laser frequency fluctuation recorded over 500~secs and \textbf{e}, Allan deviation of the fractional frequency stability. The fit follows $\tau^{0.95}$.}
    \label{fig3_closed}
\end{figure}
Figure~\ref{fig3_closed}a shows the schematic of the phase-noise measurement setup for the hybrid-integrated EOSIL laser. The laser output, at a wavelength of around 1553.8~nm, is mixed with two independent, fully stabilized optical frequency combs, and the two resulting beatnotes are recorded for phase and frequency analysis (see Methods). Figure~\ref{fig3_closed}b presents the frequency noise power spectral densities (PSDs) of the hybrid-integrated laser in three configurations: free-running, self-injection locked (SIL), and SIL with electro-optic (EO) feedback activated (EOSIL). Self-injection locking suppresses the laser frequency noise by orders of magnitude over a wide bandwidth. When the EO feedback loop is engaged, EOSIL actively corrects the laser frequency with reference to the $\mu$FP resonance, further suppressing the frequency noise below the SIL level. Using the $\beta$-separation method~\cite{di2010simple}, the integral linewidth ($\Delta\nu_{\beta}$) and fundamental Lorentzian linewidth ($\Delta\nu_0$) of the free-running DFB laser are 3.8~MHz and 81.4~kHz, which are suppressed to 38~Hz and 58~mHz, respectively. Also, as shown Fig.~\ref{fig3_closed}b, the frequency noise of the free-running and EOSIL stabilized laser at 1~Hz offset are $9.1\times10^{12} Hz^2~Hz^{-1}$ and $9.7\times10^{1} Hz^2~Hz^{-1}$, respectively which indicates about 11-orders of magnitude suppression of frequency noise PSD thanks to the EO active feedback. Figure~\ref{fig3_closed}c shows the radio-frequency power spectrum of one beatnote ($\Delta f_2$) before and after EOSIL is applied, which clearly confirms the linewidth suppression achieved with the proposed architecture. Figure~\ref{fig3_closed}d shows the measured laser frequency fluctuations when the EOSIL feedback is applied, with an absolute frequency drift of about $155~Hz~sec^{-1}$ which corresponds to the slow $\mu$FP cavity resonance drift~\cite{cheng2025harnessing}. Figure~\ref{fig3_closed}e shows the Allan deviation (ADEV) of the laser's fractional frequency stability, reaching $8.68\times10^{-13}$ at 1~s and a minimum of $1.1\times10^{-13}$ at a 52~ms averaging time.

\section*{Discussion}
We have demonstrated a hybrid-integrated laser that combines self-injection locking with modulation-free electro-optic stabilization on a single foundry-fabricated silicon photonic chip, referenced to a compact vacuum-gap micro Fabry-P\'erot ($\mu$FP) cavity. The EOSIL architecture reduces the integral and fundamental linewidths of a commercially available DFB laser from 3.8~MHz and 81.4~kHz to 38~Hz and 58~mHz, respectively, and reaches a fractional frequency instability of $8.68\times10^{-13}$ at 1~s. The EOSIL architecture proposed and demonstrated here enables a sub-Hz fundamental linewidth from a commercially available chip-integrated laser through a standard integrated silicon photonic chip, enabling portable, stable optical sources in a manufacturable, scalable, low-cost silicon photonic process.

The EOSIL architecture is based upon two complementary laser stabilization mechanisms. Self-injection locking passively suppresses the laser frequency noise over a wide bandwidth but does not necessarily guarantee active frequency locking and long-term stability~\cite{cheng2025harnessing}. The CCMZI electro-optic loop suppresses laser frequency noise at low offset frequencies. It derives a modulation-free error signal from the $\mu$FP and actively corrects the low-frequency and long-term frequency noise~\cite{idjadi2026chip}. As shown in Fig.~\ref{fig3_closed}b, SIL suppresses the broadband noise by orders of magnitude, after which the EO feedback pushes the low-offset frequency noise down toward the $\mu$FP thermal-noise limit. The EOSIL technique also offers practical advantages over established integrated stabilization schemes. Unlike PDH locking, the CCMZI discriminator is modulation-free, eliminating the phase modulator and demodulation chain and avoiding the RAM limitations~\cite{shi2018suppression} associated with it, making it a promising EO stabilization technique for integration into various photonic platforms.



Looking ahead, the EOSIL platform is well suited to further integration and scaling. Co-integrating the feedback electronics and eliminating the free-space coupling to the $\mu$FP would yield a fully packaged, turnkey ultra-stable source, while the foundry SOI process used here supports the co-integration of multiple stabilized lasers on a single chip. Such compact, low-noise lasers are a key enabling resource for optical frequency division and low-noise microwave and millimeter-wave generation, portable optical clocks, coherent optical communication and LiDAR, and field-deployable quantum and seismic sensing. By realizing sub-Hz class coherence in a standard silicon photonic process, EOSIL provides a scalable route toward bringing ultra-stable laser performance out of the laboratory and into deployable systems.

\section*{Methods}
\subsection*{Silicon photonic chip}
The entire photonic circuit is monolithically integrated in AMF’s commercially available standard SOI process, within a chip footprint of 8.0~mm $\times$ 1.25~mm. The laser couples into the chip through an on-chip EC with nominal insertion loss of about 3~dB. The TC, which uses an efficient thermal phase shifter isolated by surrounding trenches (with $P_{\pi}$ of 1.36~mW), divides the laser power arbitrarily. The two outputs of the MZI finally recombine in a 2$\times$2 MMI (0.15~dB insertion loss) and are detected by germanium BPDs with a responsivity of 0.96~A W$^{-1}$.

\subsection*{Phase noise measurement}
A small portion of the EOSIL hybrid-integrated laser output is used for phase noise and frequency counting. The laser output is divided equally between two branches. After polarization adjustment, each branch is mixed with one of two independent, fully stabilized optical frequency combs (Vescent FFC-100 and Menlo SmartComb-ULN), each of which is referenced to its own ultra-stable laser. The comb outputs are optically filtered around the wavelength of interest and amplified by EDFAs. The resulting interference signals are detected by balanced photodetectors (Thorlabs PDB470C), and the two beat notes are low-pass filtered to a 100 MHz bandwidth and recorded simultaneously for cross-correlation and frequency counting.
\subsection*{Laser and PIC packaging}
The hybrid-integrated laser was assembled on a custom aluminum housing. The PIC was affixed onto the aluminum substrate with silver epoxy and cured at 120~$^\circ$C for 5 min. The PCBs were then attached via set screws and wire-bonded to the PIC. A micro-thermoelectric cooler (TEC, TEC Microsystems 1MD02-020-03) was shimmed up using silicon spacers to be set 280~$\mu$m below the top of the PIC. The silicon spacers and the micro-TEC were affixed using thermal paste and silver epoxy, then cured at 80~$^\circ$C overnight. Separately, a DFB laser die (MACOM, MOAD-155D00B-LCT5) was soldered onto a metalized silicon spacer with tin-silver-copper (SAC) solder to achieve a total subassembly height of 280~$\mu$m. The DFB laser subassembly was then placed onto the micro-TEC, now affixed to the aluminum housing, with silver epoxy and thermal paste and aligned to the PIC’s input waveguide. A thermistor (Ohizumi, NBC2103F393F9900) was attached onto the DFB laser spacer with silver epoxy. The DFB laser subassembly was then cured at 80~$^\circ$C overnight. The DFB subassembly and micro-TEC were then wire-bonded to the PCBs. Using the on-chip photodiode, two 500~$\mu$m-diameter ball lenses (Nippon Electric Glass, MK-07 and LaSF015) were actively aligned to the gain chip and PIC waveguides with the laser biased at 50~mA and kept at 20~$^\circ$C. The lens in front of the laser chip was glued down first, followed by the one coupling into the PIC. Both lenses were affixed using UV-cured glue (Optocast, 3410-XTP-300K) and cured using a 365~nm LED.  

\newpage
\section*{Data availability}
All data supporting the findings of this study are available within the
article. Any additional requests for information can be directed to, and will be fulfilled by, the corresponding authors.

\section*{Acknowledgments}
We thank Dr. Nicolas K. Fontaine and Dr. Franklyn Quinlan for helpful technical discussions.

\section*{Funding}
We acknowledge support from Nokia Bell Labs, Yale University, Quantum CT, the Roberts Innovation Fund, and National Science Foundation QLCI award OMA-2016244 (P.T.R.).

\section*{Author contributions}
M.H.I. and H.C. conceived the idea. M.H.I. taped-out the photonic chip. H.C., B.L., and N.J. fabricated and characterized the $\mu$FP cavity. M.H.I. and B.L. conducted the experiments. M.H.I., H.C., and B.L. designed the PCB and package. T.Y.H. and K.K. packaged the laser and PIC. P.T.R. advised on the project. M.H.I wrote the paper with input from all authors. All authors reviewed the paper.  

\section*{Competing interests}
P.T.R. is a founder and shareholder of Resonance Micro Technologies Inc. The remaining authors declare no competing interests.

\section*{Additional information}
Correspondence and requests for materials should be addressed to Mohamad Hossein Idjadi or Peter T. Rakich.

\bibliography{EOSIL.bib}


\newpage

\renewcommand{\thefigure}{S\arabic{figure}}
\renewcommand{\thetable}{S\arabic{table}}
\renewcommand{\theequation}{S\arabic{equation}}
\renewcommand{\thepage}{S\arabic{page}}
\setcounter{figure}{0}
\setcounter{table}{0}
\setcounter{equation}{0}
\setcounter{page}{1} 

\end{document}